# Magneto-optic phonon resonances in magnetic topological $EuCd_2As_2$ via helical Raman spectroscopy

Jin Ho Kang[1,*], Liangbo Liang[2], Ioannis Petrides[3], Subhajit Roychowdhury[4,5], Kai-Chi Chang[1], Chandra Shekhar[4], Claudia Felser[4], Prineha Narang[3], and Chee Wei Wong[1,*]

[1]*Mesoscopic Optics and Quantum Electronics Laboratory, University of California, Los Angeles, CA 90095, USA*

[2]*Center for Nanophase Materials Sciences, Oak Ridge National Laboratory, Oak Ridge, TN 37831, USA*

[3]*Division of Physical Sciences, College of Letters and Science, University of California, Los Angeles, CA 90095, USA*

[4]*Max Planck Institute for Chemical Physics of Solids, 01187 Dresden, Germany*

[5]*Department of Chemistry, Indian Institute of Science Education and Research Bhopal, Bhopal-462 066, India*

* Corresponding authors: zse0147@ucla.edu; cheewei.wong@ucla.edu

**$EuCd_2As_2$ materials have two magnetic ordering states: antiferromagnetic (AFM) and ferromagnetic (FM) when their chemical tunability is utilized. While AFM-$EuCd_2As_2$ has a nonzero magnetoelectric response due to its symmetry breaking with spin configuration, FM-$EuCd_2As_2$ is an ideal candidate for studies of Weyl physics because of its minimum number of Weyl points with opposite chirality. In this article, we examine cryogenic low-frequency Raman spectroscopy of phonon modes in FM-$EuCd_2As_2$ crystals using circular polarization configurations, with support from density functional theory calculations, and investigate in-plane magneto-anisotropy by linear polarization configuration below the Curie temperature ($T_c$ = 26 K). We attribute the anomalous enhancements in Raman intensities below the Curie temperature are due to spin-phonon coupling. Furthermore, we see that *A*-mode peaks can be distinguished by magneto-helical Raman spectroscopy through the magneto-optic effect and that the degree of circular polarization (DCP) of 12.5 meV peak reaches 60% at 4.2 K and becomes saturated. We also examine AFM-$EuCd_2As_2$ below Néel temperature ($T_N$ = 9 K) to compare with FM-$EuCd_2As_2$, but we hardly observe spin-phonon coupling and find negligible DCP values due to almost zero net magnetization. Our results contribute to the understanding of the phonon dynamics and the interplay between topology** **and magnetism in FM-$EuCd_2As_2$, through helical light and external magnetic fields. This lays the foundation for utilizing state-of-the-art Weyl systems for applications in thermoelectrics, phononic devices, and topological quantum computing.**

Magnetic topological phases have garnered much interest since the interplay between the topological band structure and its magnetic order exhibits extraordinary physical properties including axion insulators [1, 2, 3], magnetic Weyl semimetals [4, 5, 6, 7], and quantum anomalous Hall insulators [8, 9, 10, 11]. Weyl semimetals have a unique topological phase arising from Dirac semimetals due to the breaking of inversion and/or time-reversal symmetries. Once these symmetries are broken, the degeneracy of the bands is split into linear band crossings with inverse spin chirality, termed Weyl points. Importantly, Weyl points always appear as a pair within the Brillouin zone since the net chiral charge sums to zero according to the no-go theorem by Nielsen-Ninomiya [12].

Recently several materials such as $EuCd_2As_2$ have been proposed as candidates for Weyl semimetals that have the ideal properties of the minimum number of Weyl points near the Fermi-level. A Weyl system with fewer points is likely to exhibit simpler and well-defined chiral properties, making it more suitable for novel applications. It is generally believed that broken inversion symmetry has a minimum of four Weyl points, while broken time-reversal symmetry in a magnetic system brings only a single pair of Weyl points. By chemical tunability, $EuCd_2As_2$ has two magnetic ordering states of antiferromagnetic (AFM) and ferromagnetic (FM) phase below its critical temperature [13, 14, 15, 16]. In the AFM phase, $EuCd_2As_2$ has a gap opening at the linear band crossing points because of its in-plane spin configuration, and thus AFM-$EuCd_2As_2$ has a nonzero magnetoelectric response similar to $MnBi_2Te_4$. However, FM-$EuCd_2As_2$ is an ideal candidate for Weyl semimetal because the FM-phase possesses one pair of Weyl points when the spin configuration is lying along the *c*-axis [4, 15].

Raman spectroscopy is a non-destructive optical method with heightened sensitivity to detect fine changes in the lattice dynamics with magnetic ordering in magnetic materials, via a linear or circular polarization-resolved light configuration [17, 18, 19, 20]. Here we uncover the phonon vibrations of FM-$EuCd_2As_2$ with its magnetic ordering using a customized high-precision cryogenic polarized Raman spectroscopy with 6-$cm^{-1}$ resolution. We show the spin-phonon coupling by the resonance intensities being enhanced below the Curie temperature. Furthermore, we report the magneto-optical effects of both FM- and AFM-$EuCd_2As_2$ by probing the phonon properties through helical light under magnetic-optic spectroscopy.

**Magnetic Weyl semimetal FM-$EuCd_2As_2$**

The crystal structure of $EuCd_2As_2$ in space group $P\bar{3}m1$ (no. 164) is shown in the left inset of Figure 1a and has a layered lattice compound structure where the $Cd_2As_2$ bilayers are sandwiched between triangular Eu layers. The $EuCd_2As_2$ undergoes an FM phase transition at around 26 K, with the spins aligned in-plane (red arrows). The top view of $EuCd_2As_2$ has three-fold symmetry, as detailed in Figure S1 in Supplementary Information (SI). To achieve a flat and clean surface for optical measurements, we cleave the bulk single crystal and attach them on a silicon substrate by a silver paste, thermally connected by a cold finger in the cryostat.

Figure 1a shows a Raman spectrum of the bulk FM-$EuCd_2As_2$ with both 532 nm (green) and 785 nm (red) excitation sources in a linear parallel-polarization configuration. The FM-$EuCd_2As_2$ has a total of four distinctive resonances with 532 nm excitation, located at 8.2 (I), 12.5 (II), 20.5 (III), 24.0, ($\omega_2$) meV. In addition, there is a very broad peak at 10.5 meV ($\omega_1$) and a small shoulder at 21.9 meV (IV). The right inset shows a magnification of resonance IV. The intensities of these phonon resonances with 785 nm excitation weaken because the intensity of Raman scattering is proportional to the fourth power of the laser frequency in non-resonance Raman scattering [21]. Resonances I, II, III, and IV are also clearly observed with 785 nm excitation, and in the same spectral positions with the 532 nm excitation. Resonances $\omega_1$ and $\omega_2$ are not observed in the 785 nm measurement but the excitation has a large background (see Figure S2).

From group theory, bulk $EuCd_2As_2$ belongs to the point group $D_{3d}$ and has a doubly degenerate $E_g$ symmetric Raman tensor and a non-degenerate $A_{1g}$ symmetric tensor. Figure 1b shows the phonon dispersion band structure from first-principles density functional theory (DFT) calculations. The theoretical phonon peak positions are marked with expected mode symmetry, which are located at 7.4 ($E_g^1$), 11.5 ($A_{1g}^1$), 18.8 ($A_{1g}^2$), and 20.0 meV ($E_g^2$) respectively. Each corresponds to experimental resonances I, II, III, and IV observed in Figure 1a. To experimentally verify and distinguish the $E$- and $A$-symmetry modes of the Weyl semimetal Raman resonances, we use our circular polarization methodology for identification, where the Raman tensor symmetry only has $A$-modes in the $\bar{z}(\sigma^- - \sigma^-)z$ excitation configuration and $E$-modes in the $\bar{z}(\sigma^- - \sigma^+)z$ excitation configuration (see Figure S3 and SI Section 4) [22]. From this circularly polarized analysis, we can assign resonances I and IV as $E$-modes (red), and resonances II and III as $A$- modes (blue) like in Figure 1a. The theoretical phonon peak positions are systematically underestimated but, overall, the resonance positions and their vibration symmetries are consistent with experimental data. The remaining resonances that show up in both circular polarization configurations, denoted $\omega_1$ and $\omega_2$, may be related to electronic transition from the Eu rare-earth element.

In Figure 1c, we present the dependence of the total Raman spectrum of FM-$EuCd_2As_2$ at 4.2 K on the incident linear polarization, while maintaining the relative angle of the analyzer in the parallel polarization configuration [22, 23]. The 0-degree polarization measurement is shown in Figure 1a (see Figure S4 of SI for a longer range of temperature points below and above the Curie temperature). In line with theoretical analysis of $P\overline{3}m1$, where the spin ordering is not considered, the polarization dependence remains isotropic, implying that Raman modes in the FM ordering have no in-plane magnetic anisotropy.

**Results of temperature dependent Raman**

To understand the effect of magnetic ordering on FM-$EuCd_2As_2$, we conducted temperature dependent Raman measurements from 4.2 K to 296 K. Figure 2a shows temperature dependent measurements in parallel polarization configuration. We do not observe any additional resonances below the Curie temperature ($T_c$ = 26 K), when compared to the room temperature spectrum. This indicates that the magnetic ordering does not significantly change the vibrational modes. Figure 2b displays $E$- and $A$-symmetry Raman resonance positions, relative to those at room temperature. The Raman shifts are in line with the general temperature-dependent anharmonicity model for phonon decay processes involving differing number of phonons [24]. The solid line represents the model fit, which explains our experimental data well of linear blue-shift down to 50 K and subsequent almost flat dependence at lower temperatures. Correspondingly, the Raman linewidth narrows at cryogenic temperatures as shown in Figure 2c, with the anharmonicity model capturing these dependencies well. Resonance IV ($E_g^2$) has a larger error bar because it is a shoulder peak. Below the Curie temperature, we do not observe any deviation from the anharmonic model due to spin phonon coupling. All anharmonic fitting parameters can be found in Tables S1 and S2 in SI Section 6.

On the other hand, we find that the intensities of peaks are affected by spin phonon coupling. Figure 2d shows intensities of resonances I, II, and III in the form of a scatter plot with left $y$-axis and with the shaded area depicting the experimental error. The intensities are divided by each resonance peak's room temperature value. The intensities tend to decrease as temperature decreases from room temperature, whereas the intensities start increasing around and below Curie temperature. Suzuki and Kamimura proposed a phenomenological theory for spin-dependent Raman intensities with temperature dependencies for magnetic materials [25]. This describes $I(\mathrm{T}) = |\mathbf{R}+\mathbf{M}\cdot(\langle S_iS_j\rangle/S^2)|^2$, where **R** and **M** each represents spin-independent and spin-dependent terms of the Raman tensor, and $\langle S_iS_j\rangle/S^2$ is the spin correlation function of the nearest neighbors. The reduced spin correlation function in the FM case has a unity value at $T = 0$ K and gradually decreases to zero as temperature increases until arriving at Curie temperature as the sample enters a paramagnetic phase above this temperature. Therefore, above the Curie temperature, the intensity of the Raman spectra arises solely from the spin-independent Raman tensor term, with the spin-dependent term vanishing due to the zero value of the spin correlation function. However, below the Curie temperature, the equation has a non-zero spin correlation function, and thus the spin-dependent term significantly impacts the Raman intensity. As shown in the intensities plot (Figure 2d), the intensities of resonances I, II, and III tend to decrease following decreasing temperature from the room temperature, whereas the intensities rapidly increase around the Curie temperature. This behavior can be explained by the non-zero spin-dependent correlation below the Curie temperature due to the FM phase. In addition, the increase of intensities below the Curie temperature is gradual due to the long-range order rather than the short-range Ising character [26, 27]. We further confirm the FM phase transition by DC magnetic susceptibility measurement with $B \parallel c$-axis, which shows the critical temperature of the FM- $EuCd_2As_2$ (Figure 2d, dark gray scatter, right $y$-axis) to be around 26 K. The temperature at which the rapid intensity increase is observed is well-aligned with the FM phase transition.

**Magneto-optical effect on helical Raman**

We examine the $A$-symmetry peaks by co-circularly polarized Raman spectroscopy under magnetic field to further investigate the magnetic ordering of FM-$EuCd_2As_2$. The $A_g$-mode is based on non-degenerate symmetric vibrations, which means that it is symmetric on all symmetry operations (i.e., inversion center, rotation, and mirror planes). However, when the material undergoes a change in magnetic ordering or in external magnetic field, time-reversal symmetry is broken, and thus the polarization selection rule of phonon vibrations changes. In this case, the Raman tensor has a nonzero anti-symmetric term due to the magnetic properties of a specific observed configuration [17, 18, 28].

The $A_g$ symmetry Raman tensor is shown in Equation 1a while the anti-symmetric Raman tensor of $A_g$ due to broken time-reversal symmetry in the FM state is shown in Equation 1b:

$$A_g = \begin{pmatrix} a & 0 & 0 \\ 0 & a & 0 \\ 0 & 0 & b \end{pmatrix} \quad (1a)$$

$$DA_g = \begin{pmatrix} a & -i\beta_1 M_z & -i\beta_2 M_y \\ i\beta_1 M_z & a & -i\beta_3 M_x \\ i\beta_2 M_y & i\beta_3 M_x & b \end{pmatrix} \quad (1b)$$

where $a$, $b$, $\beta_1$, $\beta_2$, and $\beta_3$ are constants, and $M_x$, $M_y$, and $M_z$ are the magnetizations along the $x$-, $y$-, and $z$-directions, respectively. In the paramagnetic state, $I_- = I_+ \propto |(\sigma_o^{\pm})^{\dagger} R(A_g)(\sigma_i^{\pm})|^2 = |a|^2$ via the Raman tensor in Equation 1a while, in the FM state, $I_- \propto |a + \beta_1 M_z|^2$ and $I_+ \propto |a - \beta_1 M_z|^2$ via the Raman tensor in Equation 1b. $M_x$ and $M_y$ components do not appear in the Raman tensor calculations because the setup configuration uses a backscattering geometry with the excitation light traveling in and out along the *c*-crystallographic orientation (i.e., the light polarizations are in the *ab* plane). In addition, we assume that $M_z$ is small or equal to zero due to the easy magnetic plane, which is the *ab* plane [13], and thus $I_- = I_+ \approx |a|^2$ even in the FM phase.

To examine this behavior, we built a magneto-Raman setup with polarization excitation under 500 mT magnetic field and cryogenic dependencies. Figure 3a shows the two *A*-symmetry resonances (i.e. II and III) at both $\bar{z}(\sigma^- - \sigma^-)z$ and $\bar{z}(\sigma^+ - \sigma^+)z$ polarizations at 4.2 K (top) and 100 K under 500 mT magnetic field (bottom). The inset schematics describe the atomic structure with its spin direction (red arrow) and the purple arrow shows the direction of applied external magnetic field. Although the sample at 4.2 K is deep into the FM state, the $I_-$ and $I_+$ have the same intensities. This further applies to the paramagnetic phase at 100 K, where the intensities are also the same. Figures 3b and 3c display two *A*-symmetry peaks under 500 mT external magnetic field at 4.2 K but with opposing directions. This external magnetic field induces a magnetization in the *z*-direction ($M_z$) and, therefore, with upward magnetic field, the intensities of the two *A*-symmetry peaks from $\bar{z}(\sigma^- - \sigma^-)z$ configuration are much larger than those from $\bar{z}(\sigma^+ - \sigma^+)z$ configuration. However, with downward magnetic field, the intensities of the two *A*-symmetry peaks are reversed because the opposite magnetic field reverses the sign of $M_z$ component. These experimental observations are consistent with the paramagnetic and FM behavior of the material. For reference, the spectra under 0-to-π linear excitation polarizations, under 500 mT at 4.2K and at room temperature, are shown in Figures S5 and S6, respectively.

To further understand the *A*-symmetry resonances (II and III) under the circularly polarization, we define the degree of circular polarization (DCP) of Raman intensities as,

$$\mathrm{DCP} = \frac{I_{-}-I_{+}}{I_{-}+I_{+}} \tag{2}$$

where $I_{\pm}$ is the Raman intensity from $\sigma^{\pm}/\sigma^{\pm}$ configurations. Measurements of resonance II ($A_{1g}^{1}$) with varying temperatures and magnitudes of magnetic fields are depicted in Figure 3d, which shows the slopes as a function of magnetic field for different temperatures. The solid lines are linearly fitted for each temperature under magnetic fields. With upward magnetic field, $I_{-}$ is stronger than $I_{+}$ due to the anti-symmetric component of $M_z$ in Equation 1b, and therefore, the material has positive DCP values. In addition, the DCP values are linearly increasing when the material is at lower temperatures and higher external magnetic fields because $M_z$ is proportional to the external magnetic field [29, 30]. The slopes below 10 K tend to decrease above 400 mT, and they seem to have a saturated DCP value above 400 mT because of magnetization saturation. This phenomenon can be observed again in the opposite magnetic field direction, where the DCP exhibits flat response at 4.2 K below -400 mT. The inset in Figure 3d shows the DCPs of magnetic dependent Raman measurements in cross-circular polarization configurations of $\bar{z}(\sigma^{-}-\sigma^{+})z$ and $\bar{z}(\sigma^{+}-\sigma^{-})z$. Because the imaginary term in the excitation and scattered light has opposite signs, $M_z$ in Raman tensor is canceled out. Thus, the intensity of each cross-circular polarization becomes the same: $I_{\bar{z}(\sigma^{-}-\sigma^{+})z} = I_{\bar{z}(\sigma^{+}-\sigma^{-})z} \propto |c|^2$ (see *E*-mode Raman tensor S1 in SI). We can see that the resulting DCP value, under cross-circular polarization, has almost flat response in terms of magnetic field. The contrasting DCP of resonance II ($A_{1g}^{2}$) is shown in Figure S7. The overall slope-per-temperature data is illustrated in Figure 3e, and the red and blue scatter dots correspond to resonances II ($A_{1g}^{1}$) and III ($A_{1g}^{2}$). Our ring magnets only apply up to 500 mT. While this does not fully align the spin directions to the *c*-axis (which requires a magnetic field $B_c > 1.6$ T [13]), this magnetic field nonetheless affects the Raman tensor matrix via morphic effects [31, 32, 33, 34]. We observed *A*-symmetry peaks with different intensity ratios which depend on the direction of circularly polarized configuration. From 4.2 K to approaching the Curie temperature, the slope for each peak reduces rapidly due to the reducing magnetization, although we can still see the tail of the slope above the Curie temperature because of the small residual magnetization in the sample. The inset in Figure 3e shows the slopes of the *E*-symmetry mode under cross-circular polarization, which is consistently flat over the same temperature range. We note that the slope of mode IV ($E_g^2$) is not plotted due to the large error bar in the peak determination due to it being a shoulder peak and having smaller intensity compared to the other resonance modes. The overall magnetic field and temperature dependent measurements of *E*-symmetry peaks are available in Figure S8 in SI Section 10.

**Results of antiferromagnetic $EuCd_2As_2$ Raman**

To compare how the FM ordering affects the material, we examine the AFM-$EuCd_2As_2$ sample by temperature dependent and magneto-Raman spectroscopy. Figure 4a illustrates the Raman spectra of the AFM sample in parallel polarization configuration at 4.2 K with 532 (green) and 785 nm (red) excitations. The position of each phonon resonance is very similar to what we measured on FM-$EuCd_2As_2$ due to the samples being isostructural. Furthermore, due to non-resonance Raman scattering, the intensity of Raman spectrum with 785 nm source is weaker than that of 532 nm. However, the resonance $\omega_2$ is still observed in 785 nm excitation measurement with AFM sample, while FM sample does not show this resonance. The top panel of Figure 4b shows Raman positions (top) as a function of temperature relative to those at room temperature, and the bottom panel shows the Raman linewidths in terms of temperature. The results follow the anharmonicity model (solid line fits) as in the case of the FM sample. In addition, the AFM-$EuCd_2As_2$ sample has a Néel temperature around 9 K [13, 35, 36] but we do not observe any abnormal intensity behavior below the critical temperature (for more information, see Figure S9). The results of our helical magneto-Raman cryogenic measurements performed on the AFM sample below and at the Néel temperature are provided in Figure 4c. We examine two *A*-symmetry resonances (i.e. $II_{AFM}$ and $III_{AFM}$) under both $\bar{z}(\sigma^- - \sigma^-)z$ and $\bar{z}(\sigma^+ - \sigma^+)z$ polarizations with 500 mT magnetic field at 4.2 K (measurement schematic shown in inset), but we cannot observe any significant DCP dependences of Raman intensities under the applied external magnetic field. The reason is that the AFM sample below the Néel temperature has an opposite magnetic ordering direction with the in-plane axis. As such, the net magnetization will be much smaller than a FM-$EuCd_2As_2$ sample, causing the anti-symmetric term $M_z$ in the Raman tensor to be zero or close to zero. The applied magnetic field of 500 mT is far below the critical field that flips all spins along the same direction; therefore, it does not introduce notable net magnetization in the AFM sample.

It follows that the term $M_z$ is still considerably smaller compared to that in the FM sample, and hence $I_-$ is approximately equal to $I_+$ in the AFM sample, leading to a zero or near-zero DCP. Figure 4d shows DCP values in terms of magnetic field, and the DCP value has an almost flat response.

## Conclusion

We examined the lattice phonon dynamics of FM- and AFM-$EuCd_2As_2$ magnetic Weyl semimetal and topological material using helical-polarized high-resolution cryogenic Raman spectroscopy. Our temperature- and polarization-dependent studies illuminate the phonon dynamics of each vibration resonance mode and the in-plane magneto–anisotropy, respectively. The positions and linewidths of the observed resonances for both samples are consistent with first-principles calculations. Importantly, the intensity of each phonon mode in FM-$EuCd_2As_2$ reduces as temperature decreases but tends to increase below the Curie temperature due to spin-phonon coupling. On the other hand, the intensities of AFM-$EuCd_2As_2$ sample have no observed spin phonon coupling even below the Néel temperature. Furthermore, once the intrinsic magnetic ordering occurs below the Curie temperature or an external magnetic field is applied along the *c*-axis crystallographic direction, the symmetric *A*-mode Raman tensor becomes anti-symmetric due to the induced magnetization, which introduces anti-symmetric off-diagonal tensor elements causing helical polarization dependency. DCP values are subsequently rapidly enhanced below the Curie temperature, before becoming saturated after 400 mT. In contrast, AFM-$EuCd_2As_2$ has a negligible DCP response because the net magnetization is zero or close to zero below the critical field ($B_c$ = 1.6 T). Our results uncover the phonon dynamics in these exotic materials hosting magnetic Weyl physics and topological properties by helical magneto-Raman spectroscopy, supporting applications such as spintronics, next-generation electronic devices, and topological quantum computing.

## Online Content

## Methods

### Crystal growth

#### FM-$EuCd_2As_2$

The high-quality single crystals of FM-$EuCd_2As_2$ are synthesized by the salt-flux procedure (equimolar mixture of KCl and NaCl). Elemental pieces of Eu, Cd, and As are purchased and mixed with NaCl and KCl inside a glove box for the synthesis. Eu, Cd, As, NaCl, and KCl are mixed with the molar ratio Eu:Cd:As:KCl:NaCl of 1:2:2:4:4. All the starting materials are loaded into an alumina crucible that is sealed in a quartz tube under vacuum. The tube is heated to 469˚C for 24 hours, dwelled for 24 hours, and then slowly heated to 597 ˚C for 15 hours. It is then dwelled for 24 hours, followed by heating to 847˚C, then dwelled for 100 hours. Finally, the tube is cooled to 500 ˚C for 300 hours and then quenched in air. Following the washing of the product with deionized water and vacuum filtering, we are able to recover plate-shaped crystals.

#### AFM-$EuCd_2As_2$

The high-quality single crystals of AFM-$EuCd_2As_2$ are synthesized by using Sn-flux. Elemental pieces of Eu, Cd, As, and Sn are mixed inside a glove box for the synthesis in the molar ratio Eu:Cd:As:Sn of 1:2:2:10. All the starting materials are loaded into an alumina crucible that is sealed in a quartz tube under vacuum. The tube is heated to 900˚C over 12 hours, dwelling there for 20 hours, and then slowly cooled to 500 ˚C over 200 h. It is then centrifuged at 500 ˚C for four minutes to remove excess Sn.

### Magnetization measurements

A Quantum Design MPMS3 instrument is used to measure the magnetization.

### Helicity-resolved magneto-Raman spectroscopy

Raman spectroscopy setup is built by a backscattering geometry configuration, and 532.16 nm green and 784.80 nm red excitation lasers are used with a 50× Mitutoyo objective lens to gather the PL signal from the sample. A λ/4 waveplate is used to generate circular polarization excitation. An excitation power of less than 500 μW is employed to avoid laser-induced heating. We use an ST-500 continuous flow cryostat to cool down the sample to 4.2 K. For the detection, we use a Horiba 1000M spectrometer with 1200 grooves/nm and a liquid-nitrogen-cooled back-illuminated deep-depletion CCD, specifically the PyLoN series from Princeton instruments, calibrated by a mercury lamp. The polarized signal is analyzed via a λ/2 waveplate and linear polarizer. Magnetic field extension kit for ST-500 allows us to apply different magnetic fields via placing different magnitudes of ring-like permanent magnets.

**DFT calculations**

To obtain lattice dynamics of FM-$EuCd_2As_2$, first-principles plane-wave density functional theory (DFT) calculations are performed using the Vienna Ab initio Simulation Package (VASP) with projector augmented wave (PAW) pseudopotentials for electron–ion interactions and the Perdew–Burke–Ernzerhof (PBE) functional for exchange–correlation interactions [37, 38]. The DFT+U method is used to consider the localized *f* electrons of Eu atoms, with the effective U parameter chosen as 5.0 eV [39]. FM magnetic ordering is considered in the calculations, which shows a large magnetic moment for the Eu element. Both atomic coordinates and lattice constants are optimized until the residual forces were below 0.001 eV/Å, with a cutoff energy of 400 eV and a gamma-centered *k*-point sampling of 18×18×10. Based on the optimized unit cell, we carry out phonon calculations using a finite-difference method via Phonopy [40]. Hellmann–Feynman forces in the 3×3×2 supercell with a *k*-point sampling of 6×6×5 are computed by VASP for both positive and negative atomic displacements (0.03 Å) and then used in Phonopy to construct the dynamical matrix, whose diagonalization provides phonon frequencies and band dispersion.

**Acknowledgements**

J.H.K and C.W.W. acknowledges support from the National Science Foundation (DGE-2125924). C.F. acknowledges funding from the Deutsche Forschungsgemeinschaft (DFG, German Research Foundation) for 5249 (QUAST), the Deutsche Forschungsgemeinschaft (DFG) under SFB1143 (project no. 247310070) and the Würzburg-Dresden Cluster of Excellence on Complexity and Topology in Quantum Matter—ct.qmat (EXC 2147, project no. 390858490). A portion of this research (first-principles phonon calculations) used resources at the Center for Nanophase Materials Sciences, which is a U.S. Department of Energy Office of Science User Facility. L.L. acknowledges computational resources of the Compute and Data Environment for Science (CADES) at the Oak Ridge National Laboratory, which is supported by the Office of Science of the U.S. Department of Energy under Contract No. DE-AC05- 00OR22725.

**Author contributions**

J. H. K. designed and led the study of the Raman experiment on FM- and AFM-$EuCd_2As_2$, including the measurements at low temperatures with magnetic fields. He also built the low-frequency magneto polarized Raman setup. J.H.K., L. L., I. P., K.-C.C., P. N., and C.W.W. analyzed the experimental data. L.L. performed the DFT calculations to support the experimental Raman data. S.R., C. S., and C. F. provided the bulk single crystals of FM- and AFM-$EuCd_2As_2$ used in this study. In addition, they did the magnetization measurements. J.H.K., L.L., I. P., and C.W.W., prepared the manuscripts. C.W.W. led and supported this research. All authors discussed the results.

**Data Availability –** The data that support the findings of this study are available from the authors upon reasonable request.

**Code Availability –** The underlying code for this study is not publicly available but may be made available to qualified researchers upon reasonable request from the corresponding authors.

**Additional information**

The authors declare no competing interests. Reprints and permission information are available online. Correspondence and material requests should be addressed to J.H.K. and C.W.W.

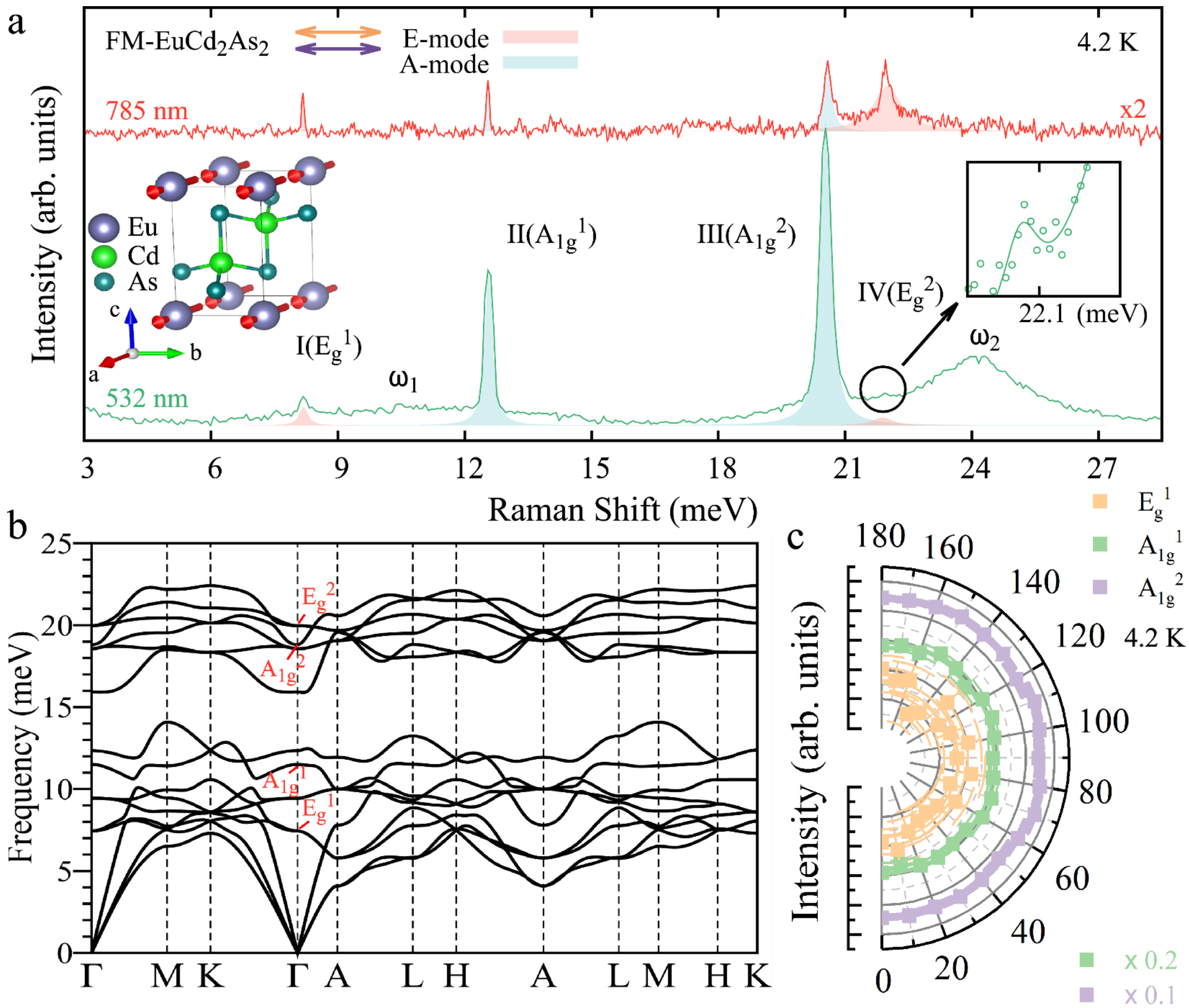

**Figure 1 | Cryogenic polarized Raman spectroscopy of magnetic Weyl semimetal FM-$EuCd_2As_2$, distinguishing *E*-mode and *A*-mode phonons. a**, Raman spectra of FM-$EuCd_2As_2$ in *E*-mode (red) and *A*-mode (blue), measured at 4.2 K in linear parallel-polarization by 532 nm (green) and 785 nm (red) excitation sources. All resonances are fitted by Lorentzian functions. The left inset shows the FM-$EuCd_2As_2$ crystal structure and the right inset is a magnification of resonance IV ($E_g^2$). **b**, Phonon dispersion band structure of FM-$EuCd_2As_2$ obtained by first-principles DFT calculations. **c**, Parallel-polarization dependence of Raman spectrum signal at 4.2 K.

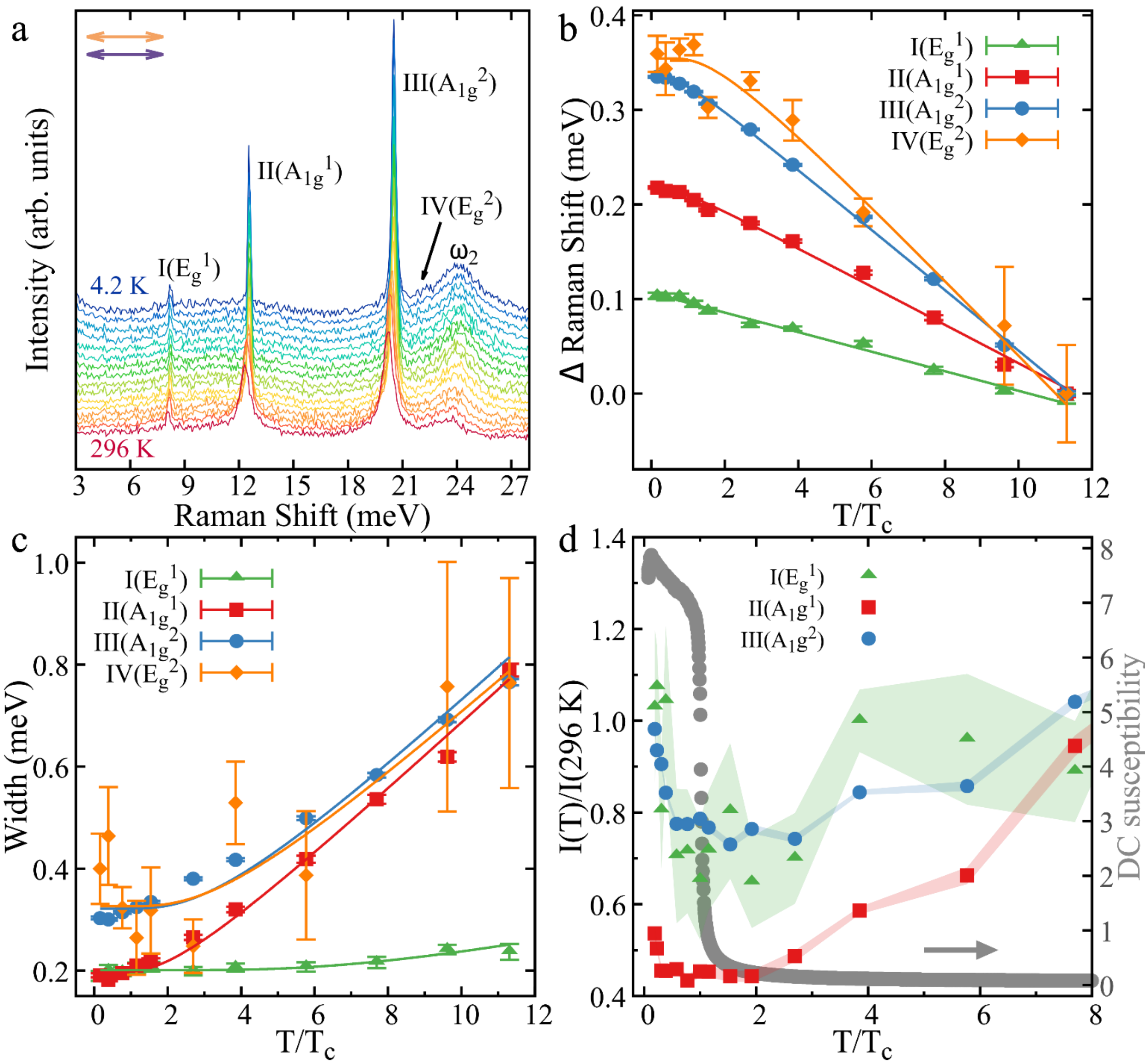

**Figure 2 | Temperature dependencies and spin-phonon coupling of FM-$EuCd_2As_2$ Raman spectroscopy peaks**. **a**, Raman spectra measured in temperatures ranging from 4.2 K to 296 K with parallel-polarization configuration. **b**, Raman shift of resonance positions I ($E_g^1$), II ($A_{1g}^1$), III ($A_{1g}^2$), and IV ($E_g^2$), relative to those measured at room temperature. Solid lines show fitting by anharmonicity model. **c**, Widths of Raman resonance I ($E_g^1$), III ($A_{1g}^1$), IV ($A_{1g}^2$), and V ($E_g^2$) extracted from Lorentzian fits. Solid lines represent fitting from the anharmonicity model. **d**, Resonance intensities of I ($E_g^1$), II ($A_{1g}^1$), and III ($A_{1g}^2$) (green, red and blue scatter points; left axis), relative to those measured at room temperature. The DC magnetic susceptibility (B || c-axis) of the FM-$EuCd_2As_2$ is shown in the dark gray scatter, on the right axis.

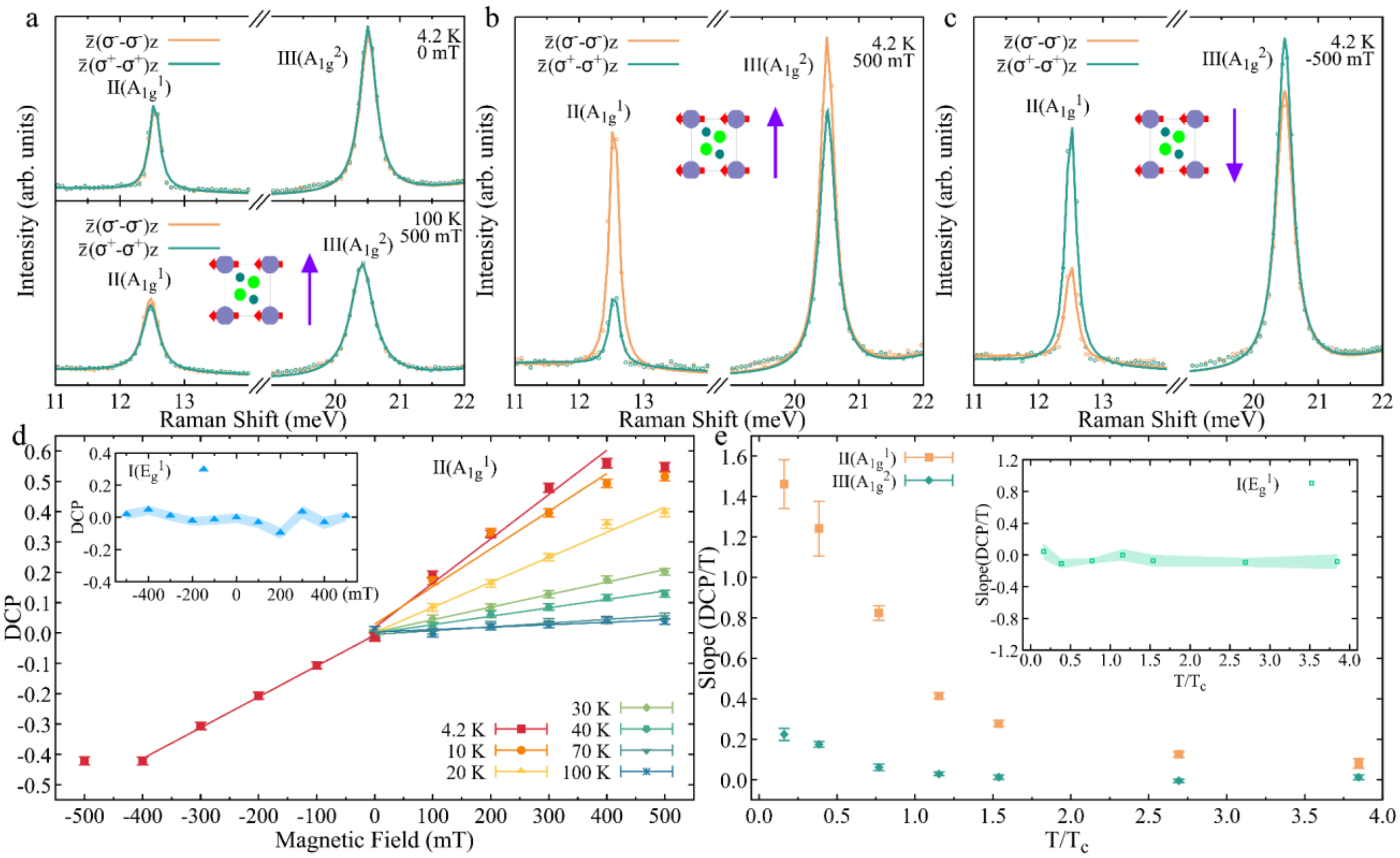

**Figure 3 | Helicity-resolved Raman spectra of the $A_g$-mode in magnetic field. a-c**, Raman spectra in co-circular polarized configurations $\sigma^-$-$\sigma^-$ (orange) and $\sigma^+$-$\sigma^+$ (green). (**a,** top) in 0 mT magnetic field at 4.2 K. (**a,** bottom) in +500 mT magnetic field at 100 K. (**b**) in +500 mT magnetic field at 4.2 K. (**c**) in -500 mT magnetic field at 4.2 K. The purple arrows in the insets show external magnetic field direction. **d**, Magnetic field dependence of the degree of circular polarization (DCP) for II ($A_{1g}^1$) from 4.2 K to 100 K. Solid lines represent linear fits. Inset represents the degree of circular polarization of $E$-symmetry resonances in cross-circular polarization. **e**, Fitted slopes versus temperature of resonances II ($A_{1g}^1$; orange) and III ($A_{1g}^2$; green). Inset shows the fitted slopes of resonance I ($E_g^1$).

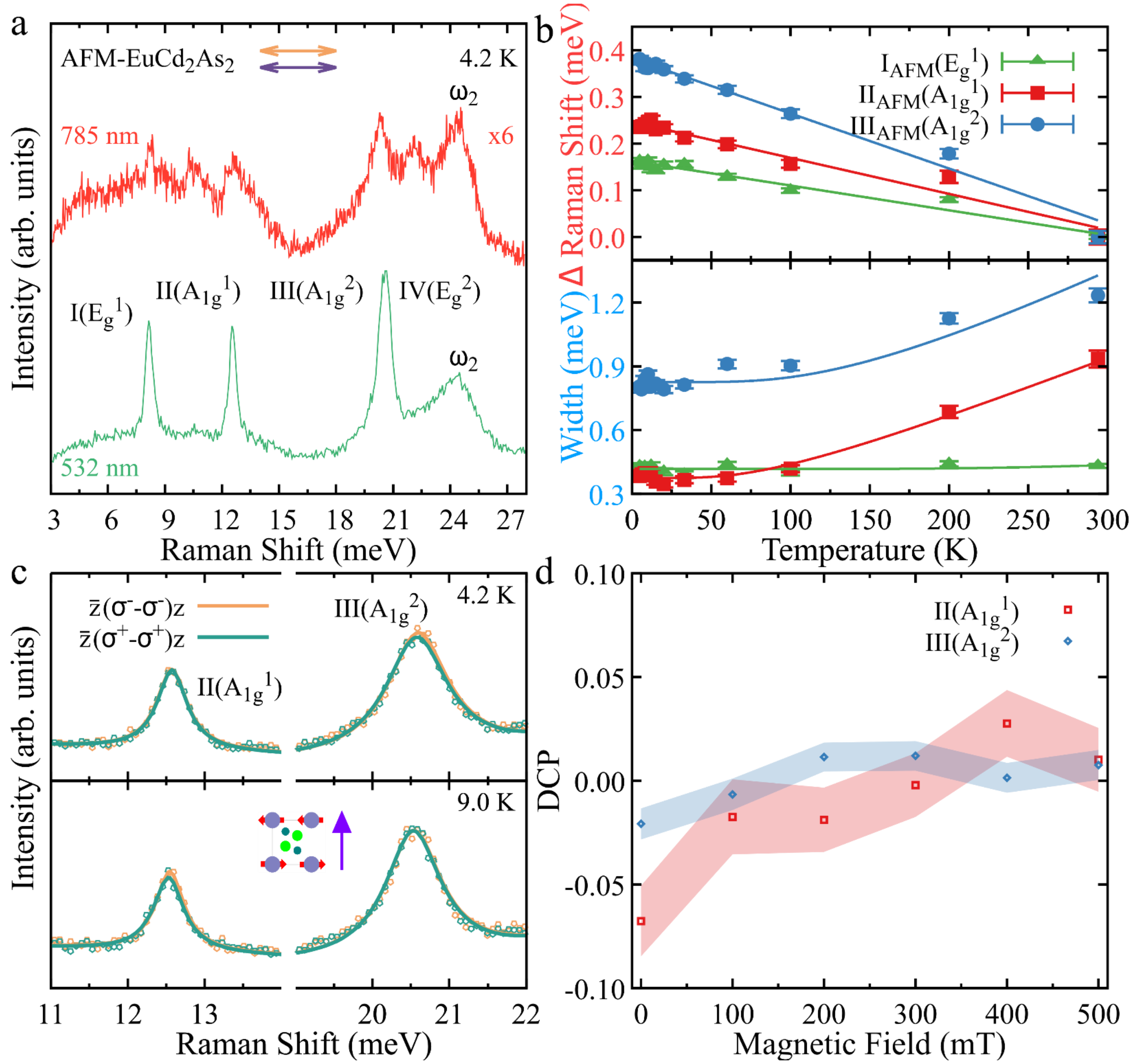

**Figure 4 | Helical magneto-Raman spectroscopy of AFM-$EuCd_2As_2$. a**, Raman spectra of AFM-$EuCd_2As_2$ in *E*-modes and *A*-modes, measured at 4.2 K in linear parallel-polarization by 532 nm (green) and 785 nm (red) excitation sources. **b**, Raman shifts (top) and their widths (bottom) for resonances I ($E_g^1$), II ($A_{1g}^1$), and III ($A_{1g}^2$). **c**, Raman spectra in co- circular polarized configurations $\sigma^-$-$\sigma^-$ (orange) and $\sigma^+$-$\sigma^+$ (green) in 500 mT at 4.2 K (top) and at 9.0 K (bottom). **d**, Degree of circular polarization of resonances II ($A_{1g}^1$) and III ($A_{1g}^2$) versus applied external magnetic field.

Supplementary Information

# Magneto-optic phonon resonances in magnetic topological $EuCd_2As_2$ via helical Raman spectroscopy

Jin Ho Kang[1,*], Liangbo Liang[2], Ioannis Petrides[3], Subhajit Roychowdhury[4,5], Kai-Chi Chang[1], Chandra Shekhar[4], Claudia Felser[4], Prineha Narang[3], and Chee Wei Wong[1,*]

[1]*Fang Lu Mesoscopic Optics and Quantum Electronics Laboratory, University of California, Los Angeles, CA 90095, USA*

[2]*Center for Nanophase Materials Sciences, Oak Ridge National Laboratory, Oak Ridge, TN 37831, USA*

[3]*Division of Physical Sciences, College of Letters and Science, University of California, Los Angeles, CA 90095, USA*

[4]*Max Planck Institute for Chemical Physics of Solids, 01187 Dresden, Germany*

[5]*Department of Chemistry, Indian Institute of Science Education and Research Bhopal, Bhopal-462 066, India*

* Corresponding authors: zse0147@ucla.edu; cheewei.wong@ucla.edu

**Contents:**

1. **Top view of crystal structure of $EuCd_2As_2$.**

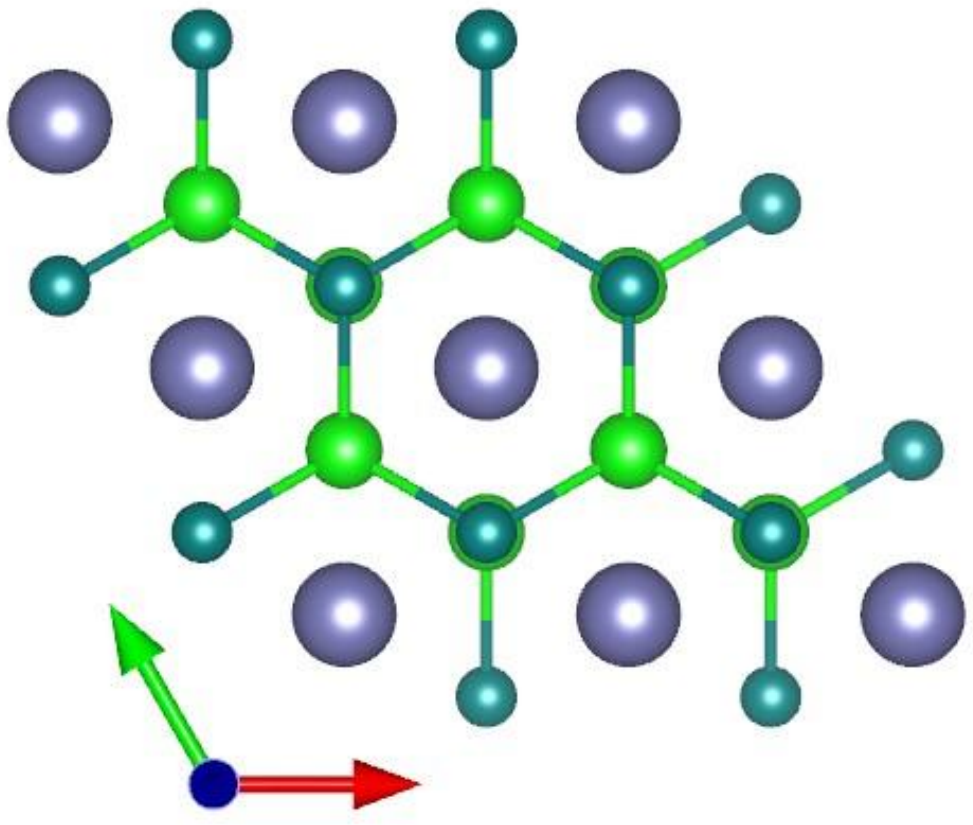

*Figure S1*. Top view of crystal structure of $EuCd_2As_2$.

2. **Raw spectra with 532 and 785 nm excitations of FM-$EuCd_2As_2$.**

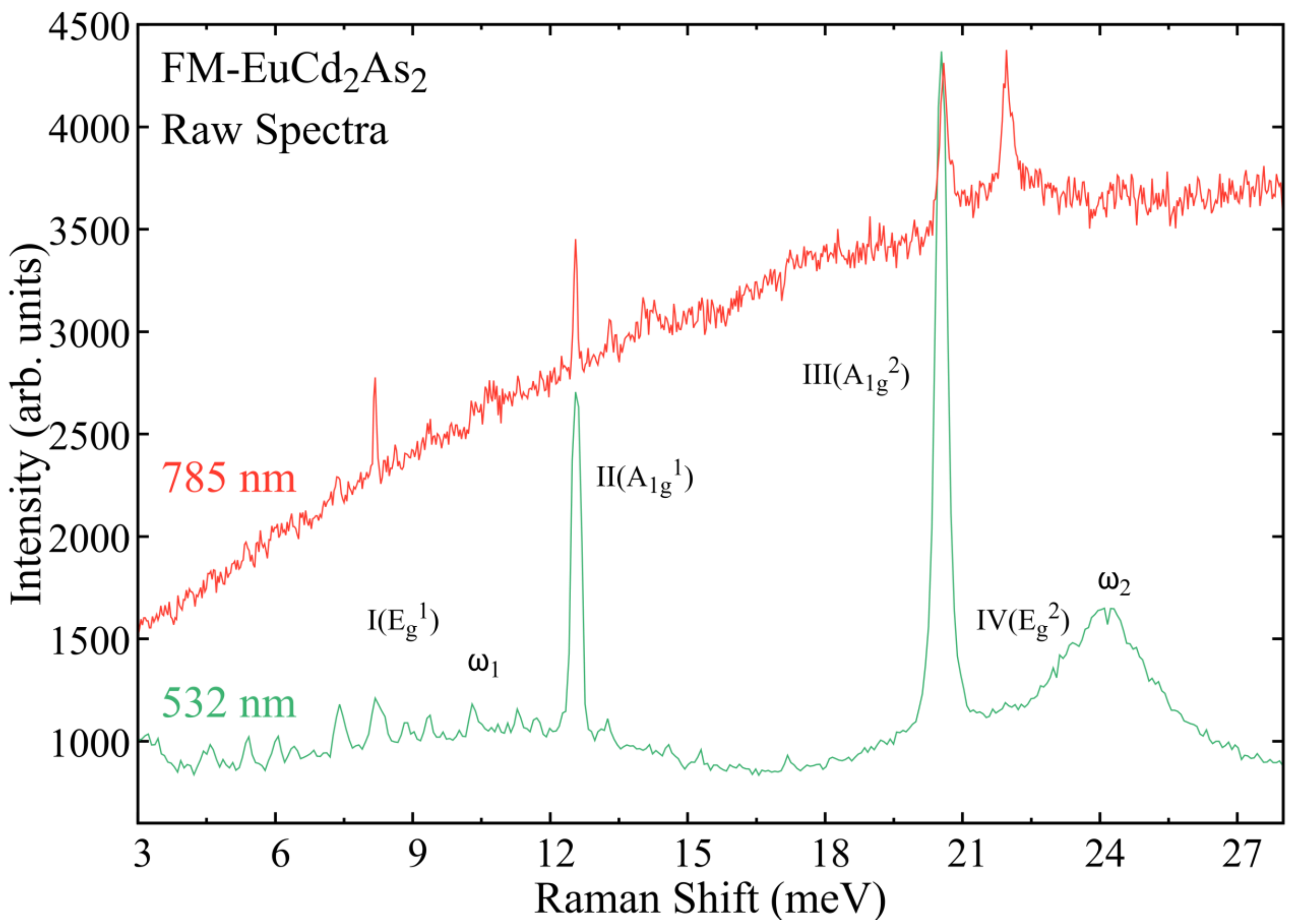

*Figure S2*. Raw spectra with 532 and 785 nm excitation sources of FM-$EuCd_2As_2$. 785 nm measurement has a stronger background signal than 532 nm spectrum.

## 3. Circularly polarized optical Raman spectroscopy of FM-$EuCd_2As_2$.

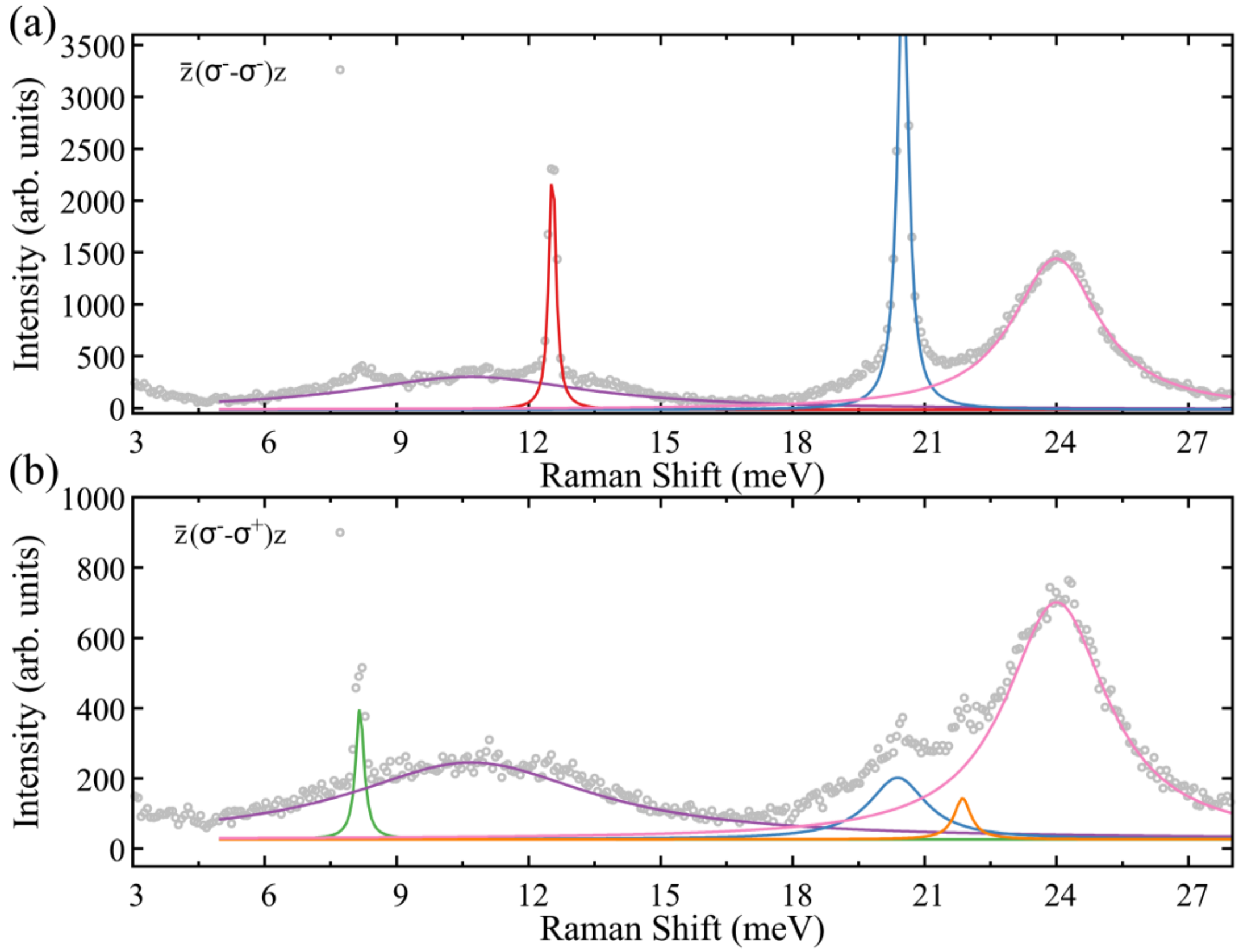

*Figure S3*. Circularly polarized optical Raman spectroscopy of magnetic Weyl semimetal of *E*-symmetry $[\bar{z}(\sigma^- - \sigma^+)z]$ (a) and *A*- symmetry $[\bar{z}(\sigma^- - \sigma^-)z]$ (b) phonons at 4.2 K.

## 4. Raman group theory analysis for $EuCd_2As_2$

The crystal structure of $EuCd_2As_2$ is in the space group $P\bar{3}m1$ (no. 164), which in turn belongs to the points group $D_{3d}$. This points group has doubly degenerate $E_g$ symmetric Raman tensor and one non-degenerate $A_{1g}$ Raman tensor [1-3].

$$R(E_g)=\begin{pmatrix} c & 0 & 0 \\ 0 & -c & d \\ 0 & d & 0 \end{pmatrix} \text{ or } \begin{pmatrix} 0 & c & d \\ c & 0 & 0 \\ d & 0 & 0 \end{pmatrix} \quad \text{(S1)}$$

$$R(A_{1g})=\begin{pmatrix} a & 0 & 0 \\ 0 & a & 0 \\ 0 & 0 & b \end{pmatrix} \quad \text{(S2)}$$

**Linear Polarization**

According to backscattering geometry, the laser travels in and out the *z*-direction, and the electrical polarization vectors of excitation and scattered light $\varepsilon_i$ and $\varepsilon_o$ lie on the *x*-*y* plane. Thus, $\varepsilon_i$ and $\varepsilon_o$ are defined by $\varepsilon_i = (\cos\theta, \sin\theta, 0)$ and $\varepsilon_o = (\cos\varphi, \sin\varphi, 0)$

$$I_A \propto \sum \left| \langle \varepsilon_o^\dagger | R(A_{1g}) | \varepsilon_i \rangle \right|^2 = |a(cos\theta cos\varphi + sin\theta sin\varphi)|^2 = |a|^2 \cos^2(\theta - \varphi)$$

$$I_E \propto \sum \left| \langle \varepsilon_o^\dagger | R(E_g) | \varepsilon_i \rangle \right|^2 = |c|^2 [\cos^2(\theta + \varphi) + \sin^2(\theta + \varphi)] = |c|^2$$

*E*-symmetry peaks should be observed in both parallel ($\theta$-$\varphi$=0) and cross ($\theta$-$\varphi$=$\pi$/2) polarization configurations independent of the polarization angle. However, *A*-symmetry peaks should only be observed in the prallel polarization because the cosin term vanishes by the cross polarization condition.

**Circular polarization**

For the co-circular polarization, incident and outgoing polarization light can be defined as $\sigma_i = \sigma_o = \frac{1}{\sqrt{2}} \begin{bmatrix} 1 \\ i \\ 0 \end{bmatrix}$.

$$I_A \propto \sum \left| \langle \sigma_o^\dagger | R(A_{1g}) | \sigma_i \rangle \right|^2 = |a|^2$$

$$I_E \propto \sum \left| \langle \sigma_o^\dagger | R(E_g) | \sigma_i \rangle \right|^2 = 0$$

For the cross-circular polarization, incident and outgoing polarization light can be defined as $\sigma_i = \frac{1}{\sqrt{2}} \begin{bmatrix} 1 \\ i \\ 0 \end{bmatrix}$ and $\sigma_o = \frac{1}{\sqrt{2}} \begin{bmatrix} 1 \\ -i \\ 0 \end{bmatrix}$

$$I_A \propto \sum \left| \langle \sigma_o^\dagger | R(A_{1g}) | \sigma_i \rangle \right|^2 = 0$$

$$I_E \propto \sum \left| \langle \sigma_o^\dagger | R(E_g) | \sigma_i \rangle \right|^2 = |c|^2$$

The circularly polarization method should be able to separate the *E*- and *A*-symmetry peaks.

## 5. Linear parallel polarization dependence of Raman spectroscopy of FM-$EuCd_2As_2$

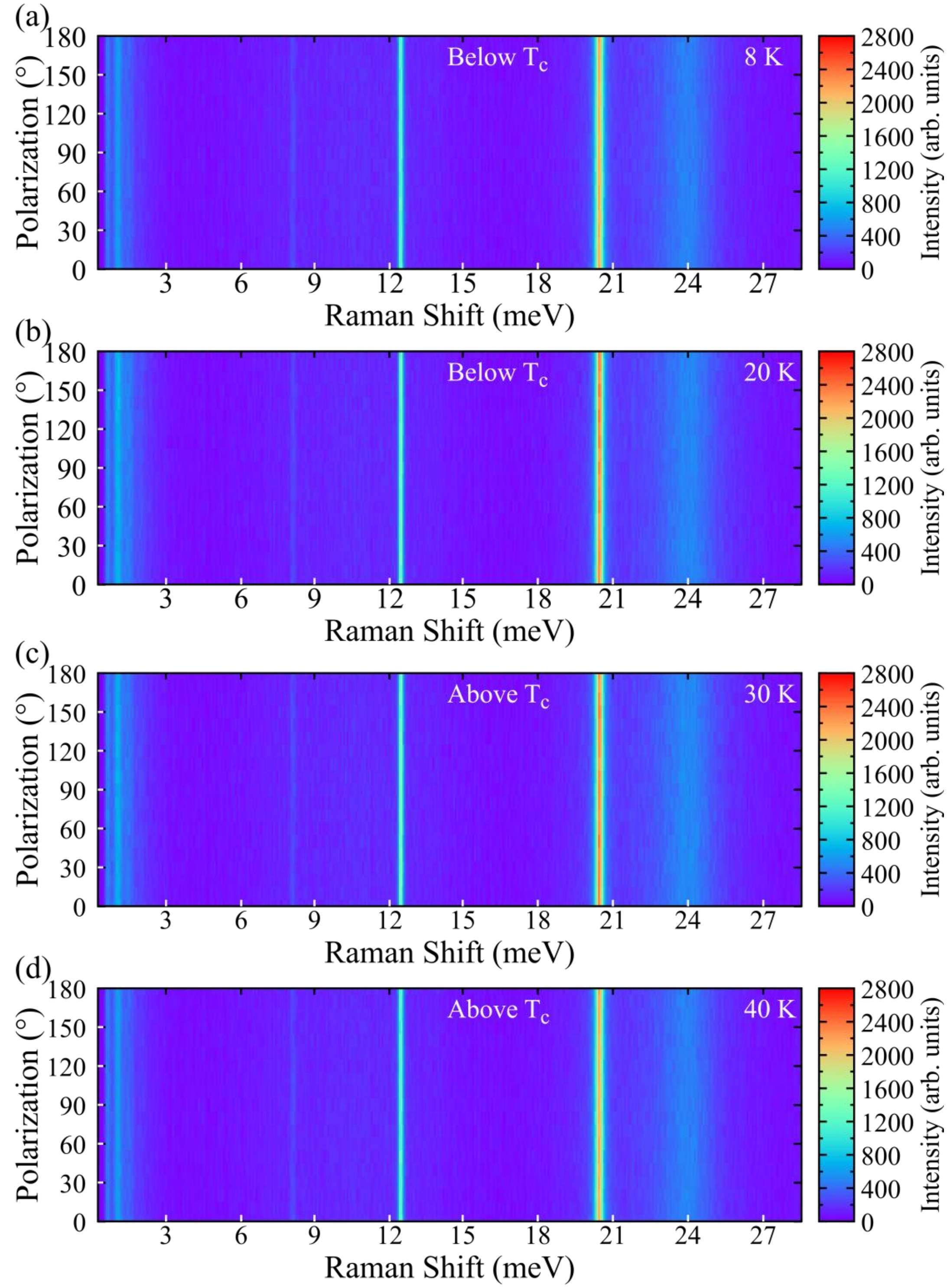

*Figure S4.* Linear parallel polarization dependence of Raman signal below Curie-temperature at 8 K (a) and 20 K (b) and above Curie-temperature at 30 K (c) and 40 K (d).

## 6. Temperature dependent anharmonic model fitting parameters.

| Peak Position | Peak I ($E_g^1$) | Peak III ($A_{1g}^1$) | Peak IV ($A_{1g}^2$) | Peak V ($E_g^2$) |
|---|---|---|---|---|
| $\omega_0$ | 0.11 ± 0.01 | 0.247 ± 0.006 | 0.368 ± 0.004 | 0.45 ± 0.08 |
| C | -0.02 ± 0.01 | -0.021 ± 0.008 | -0.034 ± 0.005 | -0.10 ± 0.09 |
| x | 0.14 ± 0.07 | 0.08 ± 0.03 | 0.08 ± 0.01 | 0.2 ± 0.1 |

*Table S1*. Fitting parameters for the peak positions by anharmonicity model given by,

$$\Omega(T) = \omega_0 + C\left(1 + \frac{2}{e^{x/T} - 1}\right), \text{where } x = T_c \frac{\hbar\omega_0}{2k_B} \quad \text{(S3)}$$

| Linewidth | Peak I ($E_g^1$) | Peak III ($A_{1g}^1$) | Peak IV ($A_{1g}^2$) | Peak V ($E_g^2$) |
|---|---|---|---|---|
| $\Gamma(0)$ | 0.201 ± 0.002 | 0.197 ± 0.005 | 0.322 ± 0.008 | 0.33 ± 0.03 |
| x | 0.96 ± 0.06 | 0.227 ± 0.009 | 0.36 ± 0.02 | 0.4 ± 0.1 |

*Table S2*. Fitting parameters for linewidths by anharmonicity model given by,

$$\Gamma(T) = \Gamma(0)\left[1 + \frac{2}{e^{x/T} - 1}\right], \text{where } x = T_c \frac{\hbar\omega_0}{2k_B} \quad \text{(S4)}$$

## 7. Linear parallel polarization dependence cryogenic-Raman spectroscopy under magnetic field of FM-$EuCd_2As_2$

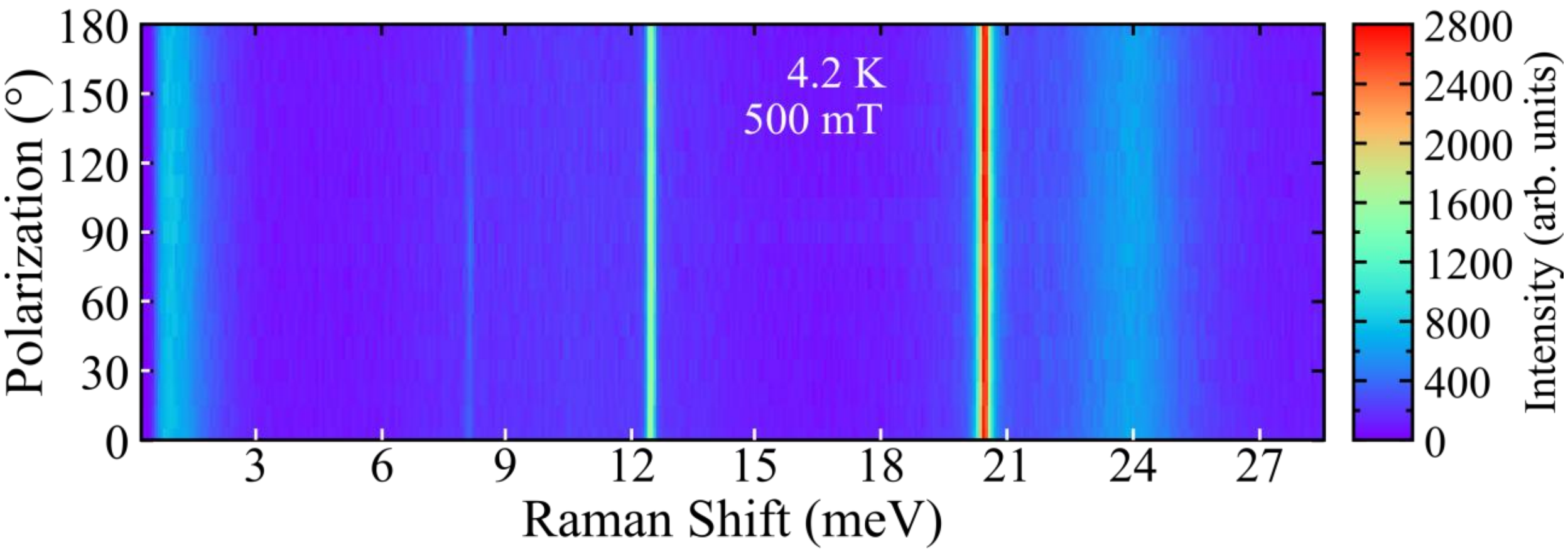

*Figure S5*. Linear parallel polarization dependence of Raman signal under 500 mT magnetic field. The magnetic field is in an upward direction.

8. **Linear parallel polarization dependence Raman spectroscopy under magnetic field of FM-$EuCd_2As_2$ at room temperature.**

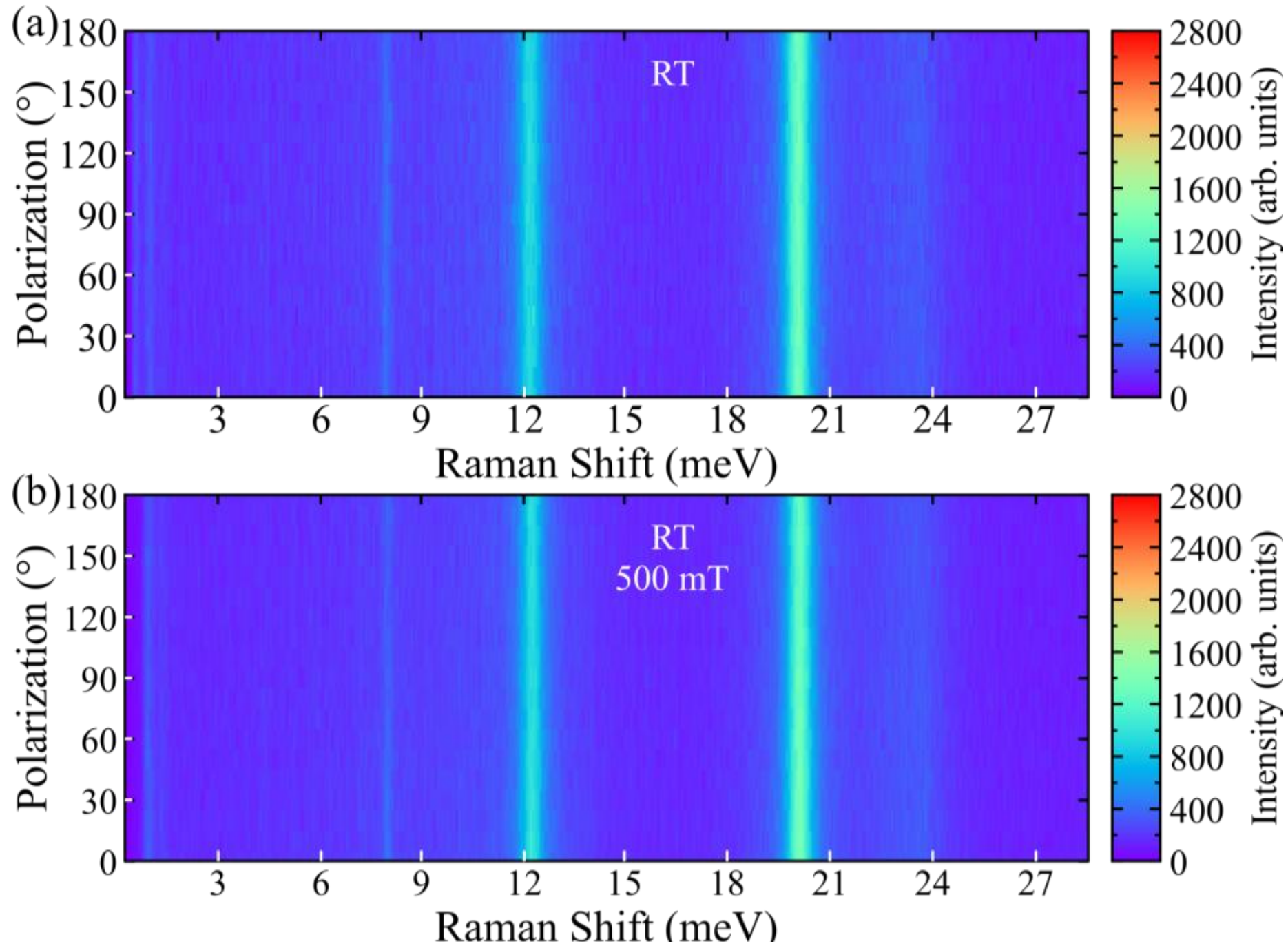

*Figure S6.* Linear parallel polarization dependence of Raman signal at room temperature (a) and under 500 mT magnetic field (b). The magnetic field is in an upward direction.

9. **Raman circular polarization of Peak III ($A_{1g}^2$) under various temperature points and magnetic fields.**

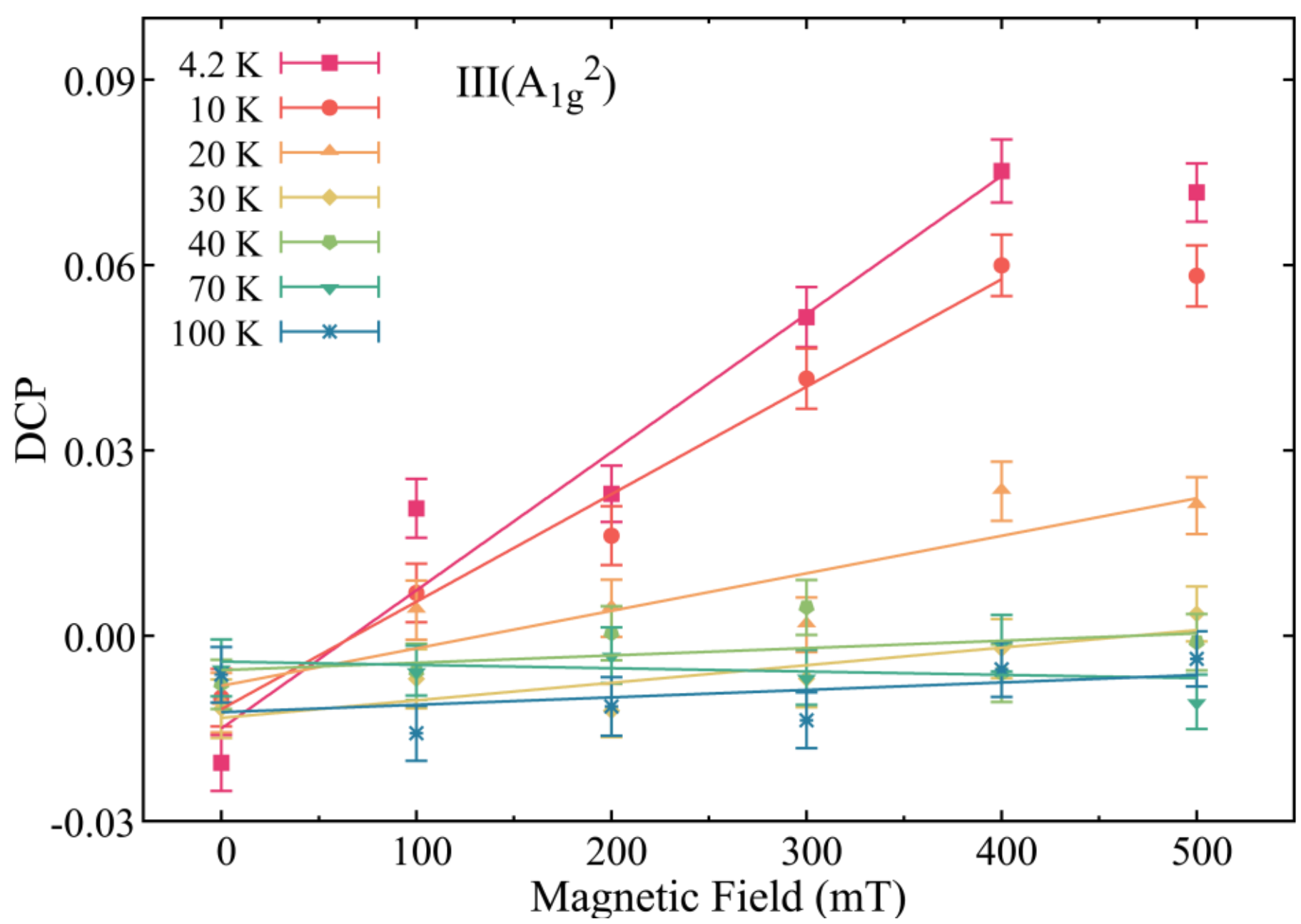

*Figure S7.* Raman circular polarization of Peak III ($A_{1g}^2$) between 4.2 K to 100 K temperature points under magnetic fields. At 4.2 and 10 K above 400 mT, the DCP values have flat response.

**10. Raman circular polarization of Peaks I ($E_g^1$) and IV ($E_g^2$) under various temperature points and magnetic fields.**

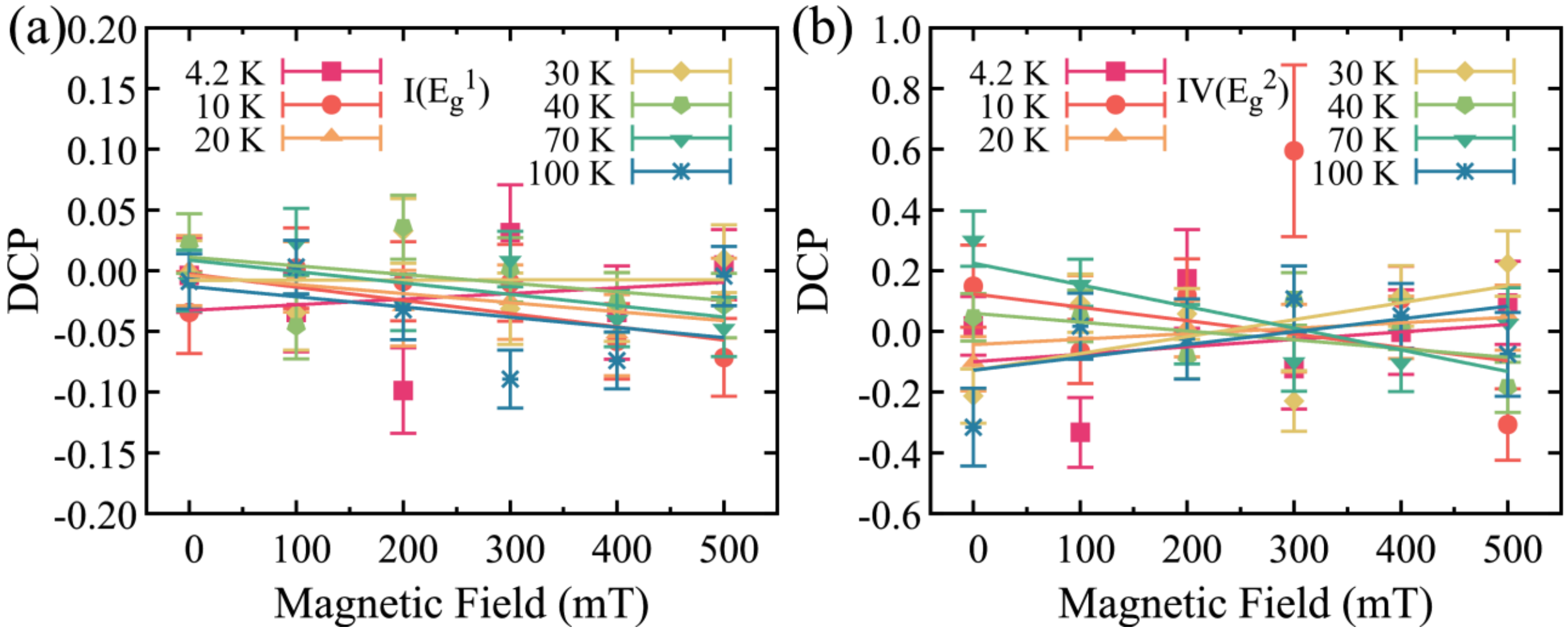

*Figure S8.* Raman circular polarization of Peaks I ($E_g^1$) and IV ($E_g^2$) between 4.2 K to 100 K temperature points under magnetic fields.

**11. Temperature dependent Raman intensities of AFM-$EuCd_2As_2$.**

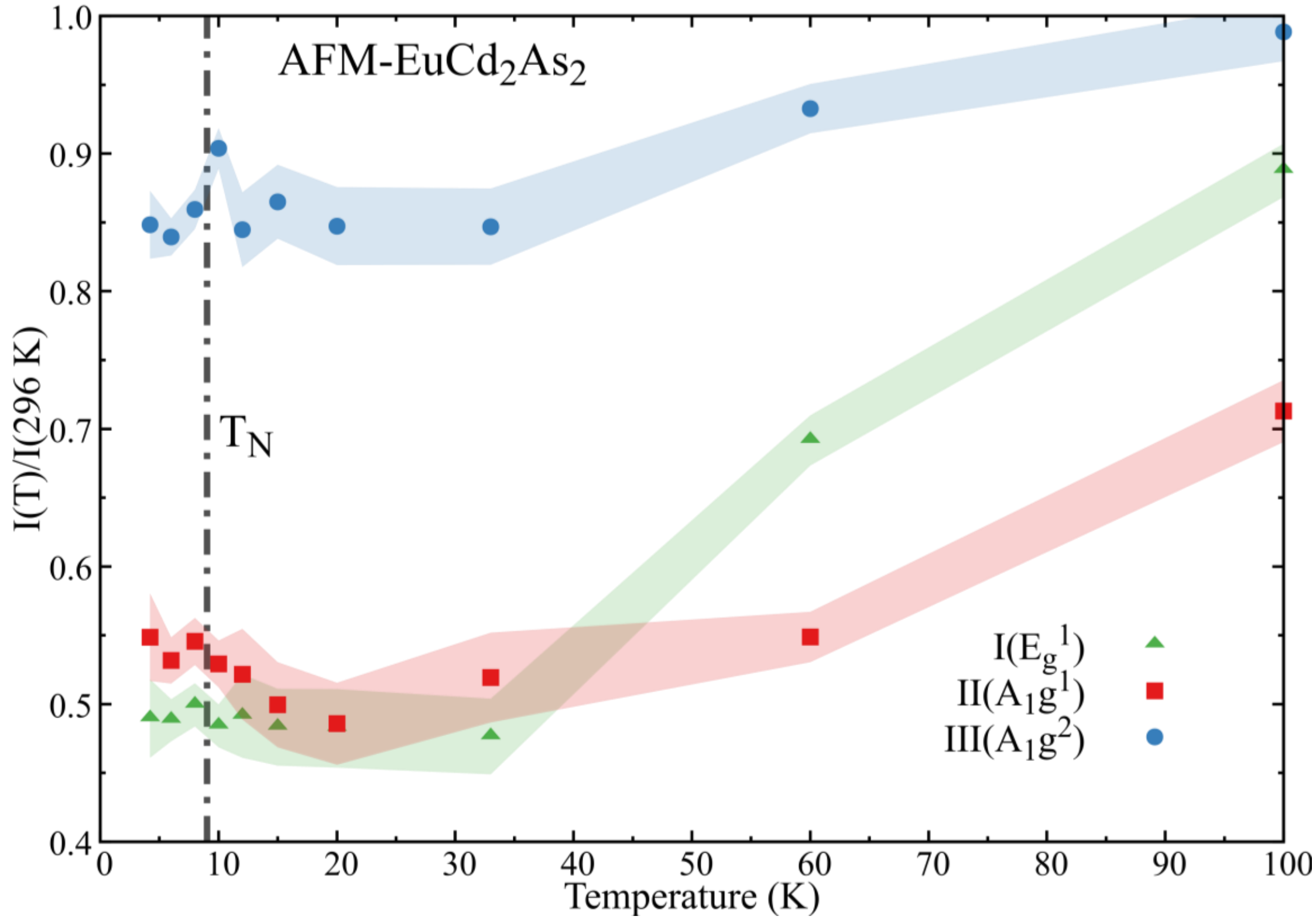

*Figure S9.* Raman intensities of temperature dependent measurement of AFM-$EuCd_2As_2$ in parallel polarization configuration. The overall intensities are divided by room temperature intensities I (296K). The vertical dashed line represents the Néel temperature of AFM-$EuCd_2As_2$ sample.

**Supporting References**